# Analysis of AODV over increased density and mobility in Intelligent Transportation System

Muhammad Ziad Nayyer
GIFT University, Department of Computer Science,
Gujranwala 52250, Punjab, Pakistan
ziadnayyer@gift.edu.pk, ziadnayyer@gmail.com

**Abstract**
Currently the area of VANET lacks in having some better designed algorithms to handle dynamic change and frequent disruption due to the high mobility of the vehicles. There are many techniques to disseminate messages across the moving vehicles but they are all highly dependent on some conditions involving flow, density and speed. The two techniques that are commonly used are AODV (Ad Hoc on Demand Distance Vector) and DSRC (Dedicated Short Range Communication). This work presents a detailed analysis of AODV. This study is focused on the use of AODV in Intelligent Transportation System. The limitations in the working of AODV routing protocol has been identified and proved. These limitations can be removed to some extent in order to increase the performance of vehicular networks and make the driving more safe and easy for a normal user as well as the implementation complications will be removed and an efficient system implementation will be possible.

## 1. INTRODUCTION

Wireless technology span has been enormous in the last decade. Now, it not only covers wireless radios or cell phones, but wide area networks, satellite equipment, point to point and point to multipoint towers. However to categorize it we can say that there are two types, one is long range such as cellular telephones, GPS, and satellite television. Second is short range wireless technology that includes wireless computer keyboards, microphones and headsets. Wireless technology is replacing wired technology due to its lack of mobility. In the current era the focus is now on globalization and coverage. Satellites are playing a vital role with help of GIS in modern world. GPS and other gadgets are becoming essential for day to day life. There is much work going on road safety, traffic and weather conditions that will aid drivers for safe driving.

Delphi Delco Electronics System and IBM Corporation introduced the concept of vehicle network and applications to aid mankind for decision making. As wireless technology grew, the area of ad hoc wireless networks got popularity and so as the vehicular network. This led to the beginning of a new era for finding new ways of C2C and C2I communication. Nearly all car manufacturers are putting their focus on installing communication, sensors and GPS equipment in vehicles to make them ready for the upcoming era of vehicular networks that will be bigger in sizes and will be denser. FCC (Federal Communication Commission) is the authority that assigns frequency and band for different wireless services [1].

Rapid growth in the field of wireless has opened many new doors. Much work has already been done and also in progress to get most of the benefit from the new technology. One focus area is traffic system. Traffic safety has always been center of gravity. Today more and more focus is on finding new ways to implement successful communication between moving vehicles.

Road safety is an important topic now days. Many transportation systems are being deployed for safe driving and those include GPS, dynamic route scheduling, and traffic conditions. There are two major network categories that are infrastructure and infrastructure less networks.

Vehicular Network, as the name shows, comprises of vehicles, also called nodes in a mobile network. Every vehicle in a VANET acts as a wireless router/node allowing other vehicles to connect and hence a network is formed. Any vehicle that goes out of the signal range is dropped out of the network.

Along with the mobile nature the node's dynamicity is also hard to maintain as a node may come and go into and out of the range and simultaneously will affect the decision concluded on the basis of the gathered data at a certain point.

VANET (Vehicular Ad Hoc Network) is a type of MANET (Mobile Ad Hoc Network). Nodes (Vehicles) communicate with each other without using a fixed infrastructure. Without central administration it becomes difficult to control several operations. Although VANET is a type of MANET but when MANET protocols are applied on VANET, it faces several challenges. There are several reasons but three main are (1) the increased node movement and rapid changes in the network, (2) MANET uses end to end connection but this depends on node density in case of VANET (3) Vehicle mobility is restricted by the road that creates many loop holes in the topology [2]. Thus there is much room for improvement that encourages researchers to put more focus on the design of protocols for vehicular networks.

Many VADD (Vehicle Assisted Data Delivery) protocols have been proposed. Some of them are L-VADD (Location First Probe). L-VADD is provided with a selected path in the forwarding direction towards the destination. It finds the closest node along that path as next hop [3]. In D-VADD (Direction First Probe), selection of direction is same as L-VADD but for an already selected direction, it chooses those nodes that are travelling in the direction of destination and moving closer to the destination node. When some certain nodes are selected a closest node is searched and selected as next hop [3].

The newly introduced protocol in the family of Ad hoc Protocols is MAODDP (Mobile Ad hoc On Demand Data Delivery Protocol) that carry out both activities one after the other including route establishment and data delivery [4]. Real Time Data Delivery (RTD) routing protocol is also used in MANET. A concept of former bandwidth reservation is used in Destination Sequences Distance Vector (DSDV) for path finding and timely data delivery. It uses multipath data delivery mechanism to accomplish data delivery of large chunks [5].

There are not many systems that would prevent a collision from happening. This need has focused on the inter-vehicle communication to speed up the decision making process through fast data transmission using dedicated short range communication and AODV. Once the data about other vehicles is received, it can be further processed to make decisions.

Intelligent Transportation System uses information and communication to effectively control and manage vehicular infrastructure. Its sole purpose is to improve vehicle safety and reduce time and fuel consumption. It is not limited to only these factors but also includes driver's ease of control and aids in decision making process. ITS uses a large variety of equipment and gadgets based on wireless sensors, GPS, wireless routers etc. The foremost and important thing is to choose an appropriate protocol and algorithm for making the communication effective between moving vehicles.

The main and foremost use of VANET applications is to disseminate safety critical messages [2], [6]. These messages need to be delivered to all vehicles. So, the two main requirements for this type of message dissemination are consistency and minimized delay. These applications make use of broadcast or multicast to disseminate packets.

## 2. RELATED WORK

The IEEE has highlighted the efforts of its members for a safer and secure traveling environment. IEEE members are contributing for the advancement of Intelligent Transportation System by introducing latest technology gadgets. They are incorporating new wireless devices and sensor in order to make it safer considering the economical factor for masses [7].

IEEE is making the use of advanced wireless communications technology and simulations in order to observe vehicular networks efficiency and working in different situations. They are making a continuous effort to make vehicles communicate successfully with other vehicles and roadside devices to gain insight of the road traffic and any upcoming blockage or natural hazard. They are also enabling vehicles to signal for emergency services in case of an emergency/accident [7].

Some of the IEEE research and development activities are mentioned below

- Automatic cruise control adaptive to traffic
- Working on mathematical models to solve problems
- Vehicular Collision Avoidance system that make use of sensors and work proactively to avoid an accident
- Real time train guiding and dispatching through wireless broadband
- Working on ITS security to make it less vulnerable to outside attacks
- Use of artificial intelligence to aid mankind for decision making [7]

2.1 Proactive vs. Reactive Approach

Two different approaches consisting of proactive approach and reactive approach is mostly used in ad hoc protocols. Proactive protocols use classical routing techniques like DSDV (Destination Sequences Distance Vector), OLSR (Optimized Link State Routing and TBRPF (Topology Broadcast Based on Reverse Path Forwarding) in which routing information about all paths is maintained whether they are used or not. There is a drawback of this technique. If the topology of the network is changing frequently then maintaining all routes could utilize a sufficient amount of network bandwidth. As the nature of vehicular network is highly dynamic, proactive approach is not suitable for these kinds of networks whereas reactive protocols such as AODV, DSR and TORA store only current routes minimizing the network load [5].

2.2 Data Dissemination Techniques

As defined in VANET architecture data dissemination between components is categorize as vehicle to infrastructure (V2I) or infrastructure to Vehicle (I2V) and vehicle to vehicle (V2V). Vehicle to Infrastructure (V2I): In V2I or I2V push based and pull based data dissemination techniques are used. In the push-based approach, the roadside unit broadcast the data to all vehicles which are in its range. Every vehicle may not be interested in the same data and that is the only disadvantage in this approach. It is suitable for applications supporting local and public-interest data such as data related to unexpected events or accidents causing congestion and safety hazards. It also generates low contentions and collisions during packet propagation. In pull based approach vehicles are enabled to query information about specific targets and responses are routed towards them. It is useful for acquiring individual specific data. It generates a lot of cross traffics including contentions and collisions during packet propagation [8].

Vehicle to Vehicle (V2V): Flooding and relaying are two approaches that can be considered for vehicle to vehicle data dissemination. In the flooding mechanism all types of data is broadcasted to neighbors. And when the neighbors receive a broadcast, it stores it and immediately rebroadcast it and hence flooding continues. It is not suitable for delay sensitive applications and also for sparsely connected or fragmented networks. This mechanism is not scalable and generates broadcast storm problem due to high message overhead during rush hours or traffic jams. In the relaying mechanism relay node is selected for disseminating the messages. The relay node is responsible for forwarding the packet further. In this approach contention is less and it is scalable for dense networks. This is due to the less number of nodes participating in forwarding message and as a result generated overhead is less [8].

3. OVERVIEW OF ITS

Intelligent Transportation System that is based on VANET provides communication infrastructure between vehicle to vehicle and vehicle to hotspot communication. The investment in this sector is increasing rapidly by transport authorities and motor car manufacturers. That time is not very far when there will be concrete standards for vehicle communication and new networks supporting new applications will be implemented.

Intelligent Transportation System uses information and communication to effectively control and manage vehicular infrastructure. Its sole purpose is to improve vehicle safety and reduce time and fuel consumption. It is not limited to only these factors but also includes driver's ease of control and aids in decision making process. ITS uses a large variety of equipments and gadgets based on wireless sensors, GPS, and wireless routers.

3.1 National ITS Architecture

National ITS Architecture is a general structure. It provides a common framework to develop multiple design approaches, each one customized to meet the individual needs, where as the benefits of common frame work remains the same. The general architecture elaborates the basic functions such as request a route, provide network information and provide information about the actual place where these functions are stored on vehicle or on Infrastructure, flow of information is V2V or V2I, and the communication requirements for the information flows between fixed points. These functions are necessary to execute to achieve a successful service. General architecture also addresses the national and international interoperability and issues like economy of scale [9].

3.2 User Services

User services document describes the users view of ITS. Almost all type of users including common user to tech users are considered. The first user services document was developed by USDOT and ITS America with significant stakeholder input and became a part of National Program Plan. User service document is to allow system or project definition to begin by establishing the high level services. These services will be further used to address identified problems and needs [9].

4. OVERVIEW OF TRAFFIC FLOW THEORY

Traffic flow theories are all about how vehicles, drivers, and the infrastructure interact with each other. The infrastructure comprise of highway system and its operational elements, together with control devices, sign boards, and marks [10].

4.1 Traffic Flow Elements

Traffic flow elements include density, lane changes per vehicle, and speed differential

**Flow:** Flow is equivalent to hourly rate at which vehicles pass through a certain point on the highway during the time period less than an hour

**Density:** Density is the number of vehicles occupying a given length of highway or lane

**Speed:** Speed is the distance traveled by a car in unit time

## 5. OVERVIEW OF AODV

AODV is a variation of DSDV (Destination Sequenced Distance Vector). The basic feature of AODV is to minimize the broadcasts. AODV only establishes route on demand and make use of destination sequence numbers to keep track of current routing paths [11].

### 5.1 Route Discovery

Current and updated entry of a route is checked in routing tables before starting transmission. If an entry is not found, AODV initiates route discovery process. A RREQ message containing its IP address, current sequence number, destination IP address, destination last sequence number and broadcast ID. Each time a source node generates a RREQ message, the broadcast ID is incremented. Sequence numbers are used to find out the timeline of each data packet, broadcast ID & IP address together form a unique identifier to uniquely identify each request. RREP packet is received in reply of a RREQ message that contains newly created route information. Source node broadcasts RREQ packets to its neighbors and sets timer to wait for a reply. In process of RREQ, the node stores the reverse path in order to reach the source node for the delivery of RREP. A life time for the reverse path is. The reverse path is deleted if not used within that period. In case RREQ is lost, the source node rebroadcasts it and reinitiates route discovery process [11].

### 5.2 Setting up of Forward Path

When a RREQ is received by destination or an intermediary node, it generates a RREP and UNICAST it towards the source using the reverse path and also stores the path into its routing table. When a RREP packet is received by source node, it means a route from source to the destination has been established [11].

### 5.3 Route Maintenance

Only routes that are currently required by the source node are stored. Source node can reinitiate the route discovery in case it loses an active path due to mobility. On the other hand if an intermediary or destination node moves out of the range then the predecessor node of the break sends Route Error (RERR) message to the affected active upstream nodes. Thus, these nodes keep forwarding the RERR messages to their predecessor nodes until the source node is reached. Upon receiving RERR message, source node either stops transmission if the route is no more required or reinitiate the route discovery if the route is still needed [11].

## 6. SIMULATION

Two simulation scenarios has been considered to evaluate the parameters under consideration i.e. density, mobility, packets received/packets lost, throughput and delay. Implementation of AODV has been done using NS2 and XGRAPH utility has been used to draw graphs.

In simulation scenario 1, 6 nodes are deployed in an 800 × 800 area. Radio propagation model is being used in this simulation with antenna type as Omni directional.

In simulation scenario 2, 9 nodes are deployed in an 800 × 800 area. Radio propagation model is being used in this simulation with antenna type as Omni directional.

### 6.1 SCENARIO 1, 6 NODES AND ONLY ONE NODE MOBIL (LESS DESITY AND MOBILITY)

Simulation scenario 1 consists of 6 nodes. Node 0 is behaving as source node and Node 5 is behaving as destination as you can see in Fig. 1. Only Node 5 shows mobility. At time 1.0ms Node 0 broadcasts a signal for node discovery and then starts data transmission using Node 2 and Node 4 as intermediate nodes.

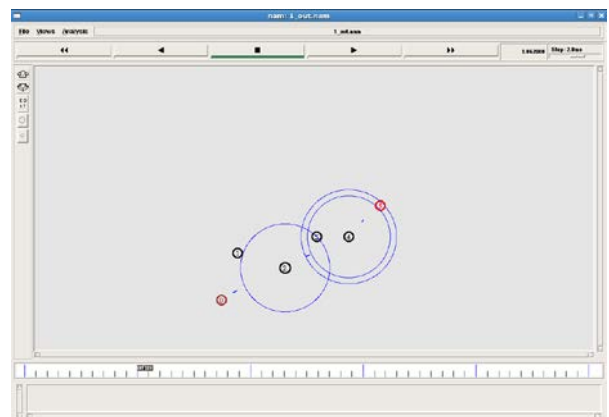

Fig.1 Data Delivery through 2 and 4

At time 2.0ms Node 5 starts moving away from Node 4 range and at time 3.0ms it goes out of the range causing the transmission to stop as you can see in Fig. 2. Due to this movement, Node 5 stops receiving packets from Node 4. Some packets are dropped and a rerouting is in progress.

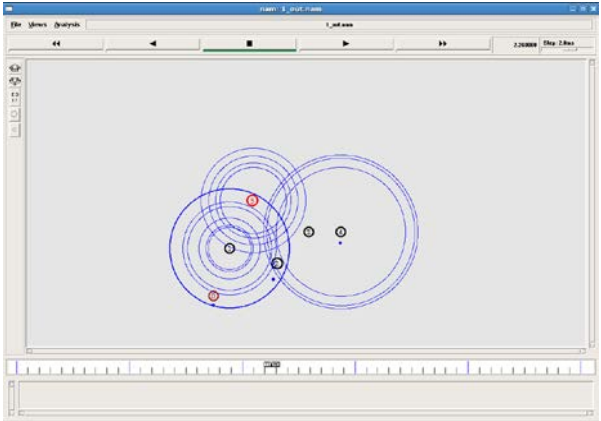
Fig.2 Node 5 Out of Range, Packets Dropped, Reroute Discovery in Progress

In Fig. 3 you can see Node 5 is now directly in reach of Node 1 so, the traffic is redirected from Node 0 to Node 5 via Node 1. The simulation ends at time 5.0 sec.

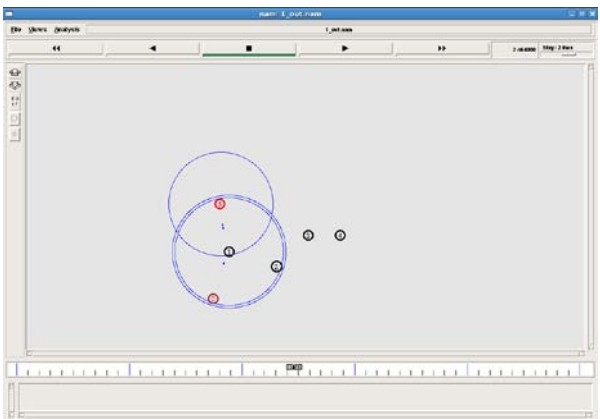
Fig.3 Data Transformation through Newly Discovered Route

## 6.2 SCENARIO 2, 10 NODES AND 4 NODES MOBIL (INCREASED DENSITY AND MOBILITY)

Simulation scenario 2 consists of 10 nodes shown in Fig. 4. Node 0 is behaving as source node and Node 5 is behaving as destination. Node 0, 7, 4 and 5 highlighted with red show mobility. At time 1.0ms Node 0 broadcasts a signal for node discovery and then starts data transmission using Node 7 and Node 3 as intermediate nodes.

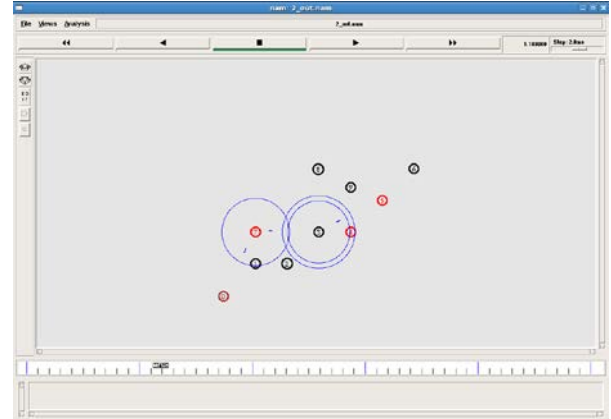
Fig.4 Data Transformation through 7 and 3

At time 1.5ms Node 5 starts moving away from Node 3 range and at time 2.3ms it goes out of the range causing the topology to break in Fig. 5. Packets are dropped by Node 3. During the movement of Node 5 before topology break Node 4 also shows mobility at 2.0ms but as this Node was not involved in data transmission it does not affect the transmission.

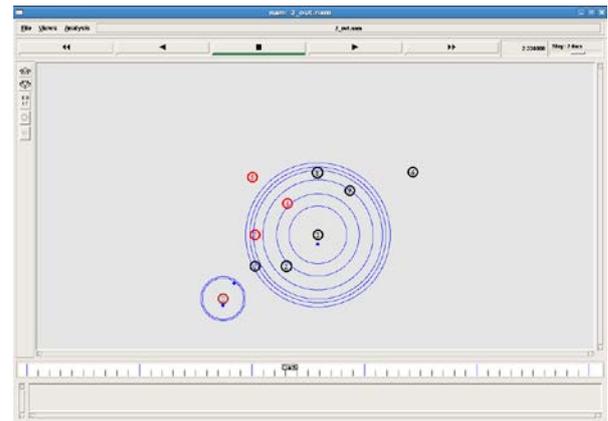
Fig.5 Node 5 Out of Range of Node 3, Packets Dropped, Reroute Discovery in Progress

In Fig. 6 through reroute discovery a new path is being found to reach Node 5 that is through Node 7.

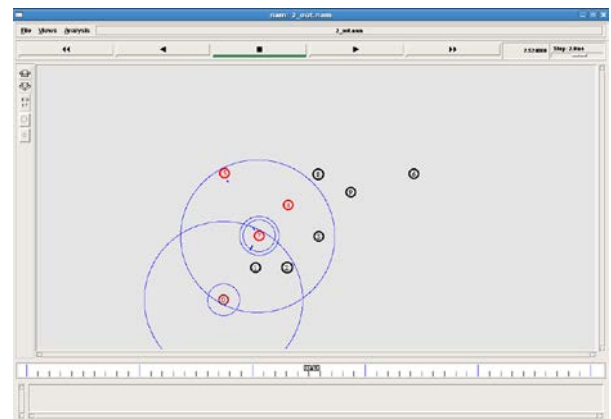
Fig.6 Data Transformation through Newly Discovered Route

At 2.5ms in Fig. 7 Node 7 starts moving away from Node 5 causing the topology to break. Packets are dropped by Node 0. Node 4 is now in reach of Node 5 so, the transmission continues through newly discovered route involving Node 1 and Node 4 as intermediate Nodes.

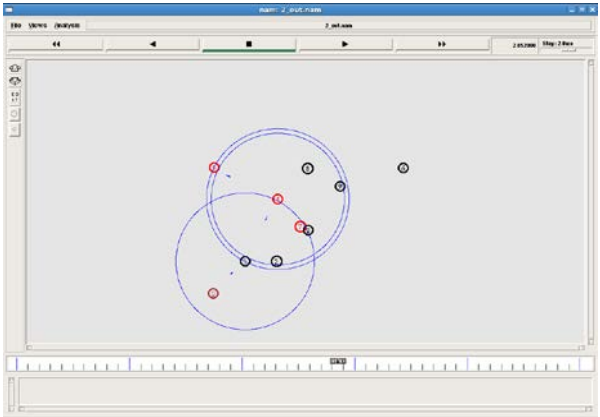

Fig.7 Data Transformation through Newly Discovered Route

In Fig.8 at time 3.0 Node 0 starts moving away from Node 1 until topology breaks.

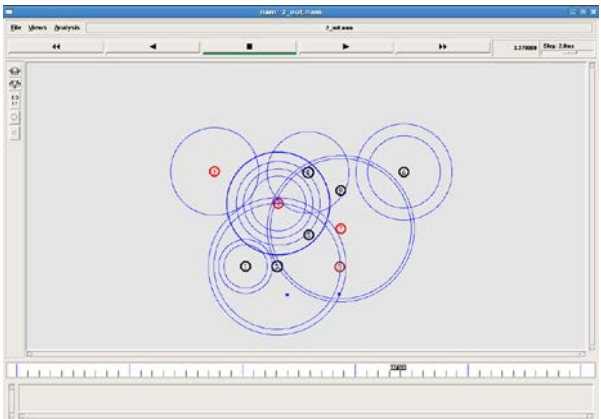

Fig.8 Node 0 Out of Range of Node 3, Packets Dropped, Reroute Discovery in Progress

In Fig. 9 some packets are dropped and a new path is being found through broadcasting and the transmission continues involving Node 9 and Node 4 to reach Node 5. The simulation ends at time 5.0 sec.

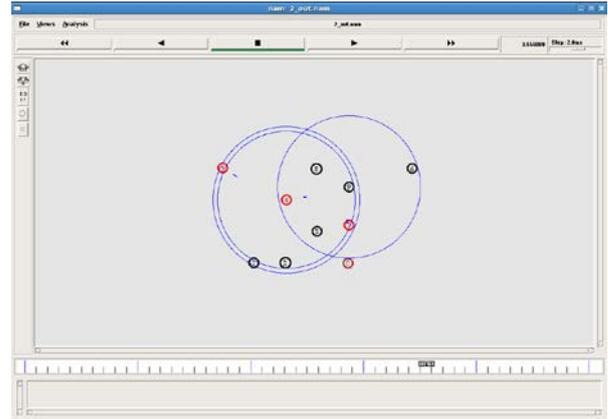

Fig.9 Data Transformation through Newly Discovered Route

## 7. ANALYSIS AND FINDINGS

An analysis has been done observing three parameters that are throughput, density and mobility and their effect after the increase in each parameter has been noted.

7.1 Evaluation Methods

There are numerous ways to study the performance of VANET protocols, however some basic methods are mentioned below

- Analysis of time complexity, delay and efficiency
- Performance comparison through simulation. Commonly used simulators are OPNET and NS2. Simulation results depend on the tool being used and parameters value being set
- Analysis of the performance using real world data but this method is not suitable for a variety of protocols [12]

The characteristics of properly designed and implemented MANET protocols are

- Quickly adapt to the runtime changes in the network
- Consume less communication and computing resources

7.2 Chosen Methodology

Qualitative and Quantitative analysis of AODV algorithm has been done using different scenarios on the basis of following parameters

- Packets Lost
- Packets Received
- Delay
- Throughput

Packets lost and packets received ratio has been studied and so as the other parameters using simulation based on NS2. XGRAPH utility has been used to draw graphs. Two scenarios with 6 and 10 nodes respectively have been taken into consideration and parameter change with the increase of density and mobility has been recorded. After separate analysis of both scenarios a combined analysis is being conducted to better focus on the changes.

7.3 Performance Metrics

Performance metrics, both qualitative and quantitative are used to assess the efficiency and advantage of a routing protocol. Some popular qualitative properties of an ad hoc routing protocol are mentioned below

*7.3.1 Qualitative Metrics*

- **Distributed:** As MANET does not support the idea of centralization, hence it requires that every protocol should be executed in a distributed fashion.

- **On Demand Operation:** Due to the dynamic nature of MANET, the routing algorithm must adjust to the ever changing traffic pattern quickly on demand, resulting into low power consumption and more efficient use of network bandwidth [13].

- **Loop Free:** For a protocol to be efficient and reliable, it must be loop free.

- **Security:** MANETs are more vulnerable to security threats as compared to Infrastructure networks so, security measures must be made like use of IPSEC to secure transmission.

- Entering/Departing Nodes: Detection of moving in and moving out nodes must be fast and well manageable in a way that there should be no need to reconstruct the whole network for a tiny change.

- **Unidirectional/Bidirectional links:** MANETs with its dynamic nature must support both kind of transmission that is unidirectional and bidirectional [13].

*7.3.2 Quantitative Metrics*

Packet Delivery Rate and Packet Delay: One main function of every routing protocol is to establish a route between source and destination. Successful delivery of data packets to the destination is another measure. An efficient protocol must have the following properties

- Higher packet delivery rate
- Minimized delay

Proactive protocols provide high packet delivery while reactive provides low [14].

- **Control Traffic Overhead:** Control overhead is low in static networks where topology and traffic stay consistent. Dynamic networks and ever changing traffic patterns increase the network overhead. Also reactive protocols introduce less control traffic overhead with static parameters while proactive protocols introduce more. If we change the parameters to dynamic from static then the situation is reversed [14].

- **Route Length:** Network overhead and delay increases when encountered with suboptimal routes. Routes those are longer than the shortest one.

- **Delivery Ratio:** Messages received over messages sent [14]

- **Transmission Efficiency:** Total number of received messages over the number of transmissions used to deliver these messages [14]

- **TCP Performance:** TCP is connection oriented and conforming protocol. It adjusts the data rate according to the information provided by the received node. In case of packets drop, TCP lowers the data rate as it assumes that the packets are dropped due to network congestion. Thus TCP is not a good choice for stress testing a MANET protocol. On the other hand we know the fact that approximately 95% of the traffic on the Internet today carries TCP. It is therefore necessary to learn how well the different routing protocols support TCP [7].

# 8. QUANTITATIVE ANALYSIS

XGRAPH has been used to conduct qualitative analysis. The parameters under consideration are packets received/ packets lost throughput and delay. The above mentioned parameters are being observed with the increase in density and mobility of traffic in VANET.

8.1 Packets Received and Packets Lost Ratio

Fig. 10 graph shows a combined analysis of packet lost and received. Red and green lines show packet lost while blue and yellow lines show packet received in scenario 1 and 2 respectively.

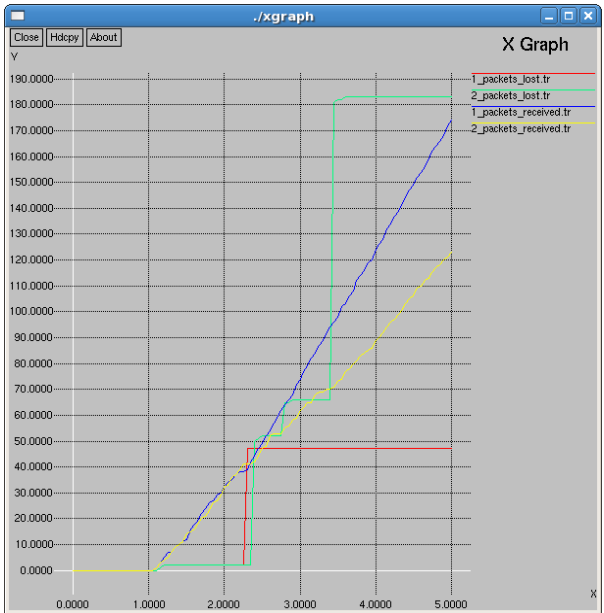

Fig.10 packets Received and Lost ratio

The graph clearly shows that with increase of mobility and density packets lost value has increased and packet received has decreased.

8.2 Throughput

Combined graph in fig. 11 with red line as Scenario 1 throughput and green line as scenario 2 throughput shows that with the increase of mobility and density, throughput has decreased.

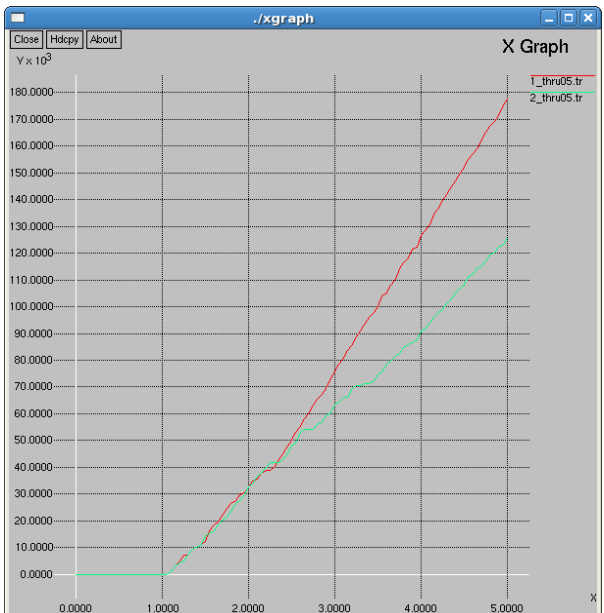

Fig.11 Throughput

8.3 Delay

Combined graph with red line in fig. 12 as Scenario 1 delay and green line as scenario 2 delay shows that with the increase of mobility and density, delay has slightly increased.

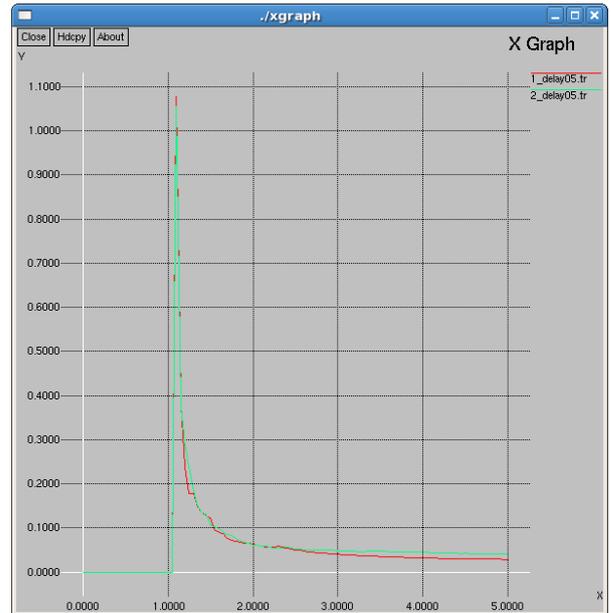

Fig.12 Delay

## 9. QUALITATIVE ANALYSIS

This section evaluates some qualitative factors of AODV. Being a reactive protocol it has some advantages that are suitable for VANETs. Among the other advantages the main two advantages are on demand rout establishment and less bandwidth consumption. It also eliminates the count to infinity problem. Other qualitative measures have been shown in table I below.

Table I
Qualitative Metrics and Findings

| Metric | Findings |
| --- | --- |
| Destination Update | Through Neighbors |
| Update period | Periodically |
| Discovery | Reactive |
| Structure | Flat |
| Route computation | Reactive |
| Multicast capability | Yes |
| Hello message requirement | Yes |
| Route metric | Fastest and shortest path |
| Unidirectional link | Yes |
| Multiple routes | Yes |
| Loop free | Yes |

## 10. COMPARISON WITH DSDV

We have compared the AODV with DSDV which is proactive in nature. We can clearly establish some facts that DSDV is tale driven and needs to keep its table up to date all the time that requires much battery power. Also a small amount of bandwidth is constantly in use in form of network overhead. A new sequence number is issued each time there is a change in network resulting into a slight amount of delay making it unsuitable for highly dynamic networks such as VANETs. Table II shows other dissimilarities between the two routing protocols.

Table II
AODV and DSDV Comparison

| DSDV broadcasts every change in the network to every node | In AODV such broadcasts are not necessary |
|---|---|
| When two neighbors enter communication range of each other<br>• This results in a network wide broadcast | If a link breakage does not affect ongoing transmission –> no global broadcast occurs |
| Similarly when two nodes drift apart from each other's range –> link breakage<br>• Also results in a network wide broadcast | Only affected nodes are informed |
| Local movements have global effects | Local movements of nodes have local effects |
|  | AODV reduces the network wide broadcasts to the extent possible |
|  | Significant reduction in control overhead as compared to DSDV |

## 11. CONCLUSION

The study shows that the AODV protocol works better under low mobility and density but the performance decreases with the increase of mobility and density. However following characteristics can be observed in the algorithm.

- Storing Only the Needed Routes
- Minimized need for broadcast
- Reduced network overhead
- Quick response to link breakage
- Use of destination sequence numbers to store loop free routes

## 12. FUTURE WORK

Intelligent transportations system can simply guide about the route and traffic or can do much more like telling in advance about the coming signal and its status also suggesting you to act to avoid stopping at the signal. It can also tell about a speedy vehicle coming from the other direction and can tell you to reduce or enhance speed to avoid a possible collision.
In future, transportation system will evolve in such a way that vehicles will not only communicate with each other but with the hotspots, satellites and other systems. Vehicles will get their position through GPS. Hotspot will provide current traffic density ahead as hotspots will also communicate with each other.

Exact timing and status of the upcoming signal will be available. Car detail information will be uploaded by inbuilt ITS for other cars and the system will advise the driver to act accordingly. Like the system will tell that currently the signal's status is this and the distance is this, if you will continue with this speed the signal will be red and if you don't want to stop you will have to increase this much speed.

Road and weather stat will also be available to ITS on the basis of which it will advise the driver to drive slow or fast. It will also guide the driver according to road condition that this road is slippery so, kindly do not speed up or apply sudden breaks.

System will also communicate with other cars in range like it will tell the driver that another car is approaching from the other direction on next crossing, if you will continue on this speed you might get collided and if you will increase or decrease your speed to this value you will avoid a possible collision.

The system will be highly adaptive and will work in two modes. One is auto mode in which it will control the car to a certain level like speed, break etc. In auto mode the car will also make maneuvers itself on the basis of gathered data like in foggy weather or in tunnel it has been observed that when an emergency break is applied there is a huge pile up as humans are unable to act that much quickly. So, when the margin is tight the manual system fails. In auto mode the car will sense the hurdle ahead and not only apply breaks but also send signal to forth coming vehicle to apply break. This message will travel in a chain making all cars to stop before colliding.

2nd is manual mode in which system will only advise the driver but driver will be purely responsible for taking action. The system will also tell the alternate, short or less dense route to reach destination.

In this study, the effect of increasing number of nodes and mobility on the network performance is under consideration. Efforts are to test the metrics with increased mobility, denser mediums and using more TCP connections. Apart from this, analysis of other routing protocols such as DSDV, DSRC and TORA and estimation of power consumption and processing costs can be done in future. Also some heavy work is needed to improve the following AODV limitations.

The limitations of AODV protocol are summarized below

- Intermediary nodes can cause conflicting routes if the source sequence number is very old and the destination sequence number of intermediary node is higher but not the most recent, thus resulting into out of date entries.

- In a result of one Route Request packet,

multiple Route Reply packets are generated that can result into serious control overhead. The periodic signaling leads to unnecessary bandwidth utilization.

- As vehicles differ in speed so, as their data rate. This can create difficulty for the evaluation of expiry time of a certain route and hence many valid routes can be expired.

- With the increase of density & mobility numerous performance metrics begin to decrease

- There is security vulnerability as AODV works on the principal that every node must cooperate with neighboring node and this can be easily exploited.

## REFERENCES

[1] R. Lind et al, .The network vehicle. A glimpse into the future of mobile multimedia, IEEE Aerosp. Electron. Syst. Mag., vol. 14, no. 9, pp. 27.32, Sep. 1999

[2] Jeppe and Lars, .Specification and Performance Evaluation of Two Zone Dissemination Protocols for Vehicular Ad-hoc Networks., Proceedings of the 39th Annual Simulation Symposium (ANSS.06), IEEE 2006

[3] Jing Zhao and Guohong Cao. VADD (2008): Vehicle-Assisted Data Delivery in Vehicular Ad Hoc Networks. Department of Computer Science & Engineering The Pennsylvania State University

[4] Humayun Bakht (2006). Implementation and Evaluation of MAODDP on Scalable Wireless Network Simulator (SWANS). School of Computing and Mathematics, John Moores University, UK

[5] Kai Han, Guanhong Pei, Hyeonjoong Cho, Binoy Ravindran, E. D.Jensenz (2007). RTRD: Real-Time and Reliable Data Delivery in Ad Hoc Networks. ECE Dept., Virginia Tech Blacksburg, VA 24061, USA The MITRE Corporation Bedford, MA 01730, USA

[6] Q. Sun and H Garcia-Molina, "Using ad-hoc inter-vehicle networks for regional alerts," Stanford University Technical Report, 2005.

[7] 13th International IEEE Conference on Intelligent Transportation Systems (ITSC 10) September 2010, in Madeira Island, Portugal

[8] Dharmendra Sutariya, Dr. S. N. Pradhan (2010), "Data Dissemination Techniques in Vehicular Ad Hoc Network", Institute of technology, Nirma University, Ahmedabad, India

[9] The US National ITS Architecture: Part 1 – Definition: Robert S. Jaffe President, Jaffe Engineering and Development Industries POB 501, Shenorock, NY 10587

[10] Wesley Ceulemans, Magd A. Wahab, Kurt De Proft and Geert Wets (July 1 - 3, 2009). Proceedings of the World Congress on Engineering 2009 Vol II WCE 2009, London, U.K.

[11] International Journal of Computing and Business Research (IJCBR) Volume 1, N. 1 (December – 2010)

[12] Liu, C., & Kaiser, J. (2005) Mobile Ad Hoc Routing Protocols Survey

[13] Subbarao, M. W. Performance of Routing Protocols for Mobile Ad-Hoc Networks. National Institute of Standards and Technology

[14] Clausen, T. H., & and Laure, P. J. (2002). Comparative Study of Routing Protocols for Mobile Ad-hoc Networks. INRIA Rocquencourt, Projet Hipercom, France

**First Author** MSCS, 2009-2011, GCU, Lahore, Punjab, Pakistan; BSCS, 2003-2007, Punjab University College of Information Technology, Lahore, Punjab, Pakistan; Currently working as lecturer Department of Computer Science, GIFT University, Gujranwala, Punjab, Pakistan; Has worked as Manager IT from 2009-2012 in GIFT University, Gujranwala, Punjab, Pakistan; Research interest is in Computer systems and networking